# Thermodynamics variables of the BTZ Black Hole with a Minimal Length and It's Efficiency

N. A. Hussein*, D. A. Eisa** and T. A. S. Ibrahim***

*Mathematics Department, Faculty of Science, Assiut University, Assiut, Egypt.
**Mathematics Department, Faculty of Science, Assiut University, New Valley, Egypt.
***Mathematics Department, Faculty of Science, Elminia University, Elminia, Egypt.

## Abstract

The aim of this paper is to study the thermodynamics variables of non-rotating BTZ black hole in 2+1 dimensional using two models; one of them is the quantum hydrogen atom model, and the other is the Collapsed Shell Model. The mass densities are expressions of the $\delta$-function. We calculate the thermodynamics variables (temperature, entropy, thermodynamics volume and heat capacity) in presence of the minimal length f. Also calculate the efficiency of holographic heat engine in absence of the minimal length for the four cases; uncharged and non-rotating, charged and non-rotating, uncharged and rotating, charged and rotating BTZ black hole respectively.

## I. Introduction

We say that the nonrenormalizability is puzzle when gravity is combined with quantum theory. Some scientist attempts to solve the problem, they have that the idea of a non-perturbative quantum gravity theory is attractive. Recently, in [1, 2] Dvali and collaborators making this idea by putting forward the "UV self-complete quantum gravity" in which the production of micro black holes is assumed to play a leading role in scattering at the Planck energy scale. In order to place a target particle more accurately the incident photon needs to have higher energy according to the Heisenberg uncertainty relation. When a new particle is creating, the photon energy is comparable to that of the particle. We can define a Planck energy, when the energy of the incident photon gets further higher, so a black holes may be create when a huge energy confined in a small scale, which called the hoop conjecture [3]. The probe meaningless happened after the formation of micro black holes, which leads to the incident photon energy, is the bigger the black hole is. As a result, the horizon radius of the extremal black hole may be deduced a minimum length.

We can define any black holes without singularity at the origin by regular black holes, the first one who research around this type of black hole is Bardeen's brilliant [4], which appeared in [5, 6] on the discussions of mass definition, casual structure, and other related topics.
In 2005, Nicolini, Smailagic, and Spallucci due to construct Schwarzschild geometry of black hole they are suggested [7] that the right hand side of Einstein's equations should be modified (energy-momentum tensor); they are obtained a self-regular Schwarzschild metric and then established new relation between the mass and the horizon radius by using Gaussian form in instead of $\delta$-function form of the point-like source and then solve the modified Einstein equations. They also want to eliminate the unfavorable divergency so they discussed the relevant thermodynamics properties and then deduced the vanishing temperature in the extreme configuration. Besides, there is a correction to the entropy in the near-extreme configuration. Since then, a lot of related researches have been carried out, such as those extending to high dimensions [8, 9], introducing the AdS background [10], quantizing the mass of black holes [11], and generalizing to other types of black holes [12, 13]. Yan-Gang Miao and Yu-MeiWu studied the thermodynamics variables of the Schwarzschild-AdS black hole with a minimal length in four dimensions [14].

In classical view of point, parameters of black hole as mass M, surface gravity κ and area A relate to the energy U, temperature T and entropy S of thermodynamics system. We are studying the 2+1 BTZ black hole because it is higher order correction to the entropy so we calculate it easy and not as uncertain. But in ground state $a \to 0$ ( a is Minimal Length )in hydrogen atom model and $l \to 0$ in other model we study the equation of state for BTZ black hole and then define classical cycles like usual thermodynamics systems. So this means that when the small BTZ black hole change to large BTZ black hole we want to the large BTZ change to small BTZ, this means that we want the system back to primary state. We study the heat engine for varies type of BTZ black hole. In this paper, we introduce the ordinary non-rotating and rotating BTZ black hole; Also we investigate equations of state and phase transitions of the self-regular BTZ black hole with two specific mass densities, where the first mass density is assumed by the correspondence to a quantum hydrogen atom and the second is a collapsed shell both of them can be regard as a point-like particle so they are two different models.

The paper is organized as follows; in section II the Hydrogen Atom Model was given. In section III the Collapsed Shell Model was given. We calculate the thermodynamics cycle and heat engine of BTZ is given in section IV. The efficiency, thermodynamics variables of static uncharged BTZ black holes and heat engine are given in section V. The efficiency, thermodynamics variables of charged and rotating BTZ black holes and heat engine are given in section VI.

## II. The Hydrogen Atom Model
### II A. The Mass Density

Using an analogy between black holes and quantum hydrogen atoms [14], we can choose the mass density of the BTZ black holes as the probability density of the ground state of hydrogen atoms:

$$\rho = M \frac{e^{\frac{-r}{a}}}{2\pi a^2} \qquad (1)$$

where *M* is the total mass of BTZ black holes and a is a minimal length.

We note that $\rho = \frac{M}{2\pi a^2}$, at $r \to 0$ which means that the origin is no longer singular.

Then the distribution function of BTZ black holes takes the form

$$\mu(r) = \int_0^r \rho(r') 2\pi r' dr' = M\left[1 - e^{\frac{-r}{a}}\left(1 + \frac{r}{a}\right)\right] \quad (2)$$

We note that at $a \to 0$, the distribution function goes to the mass of the BTZ black hole

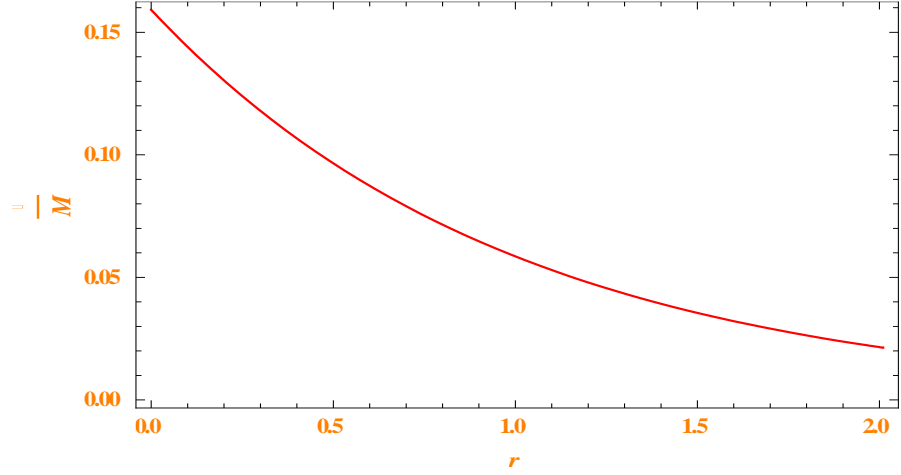

Figure 1: The relation between the mass density and the radial coordinate r (equation (1))

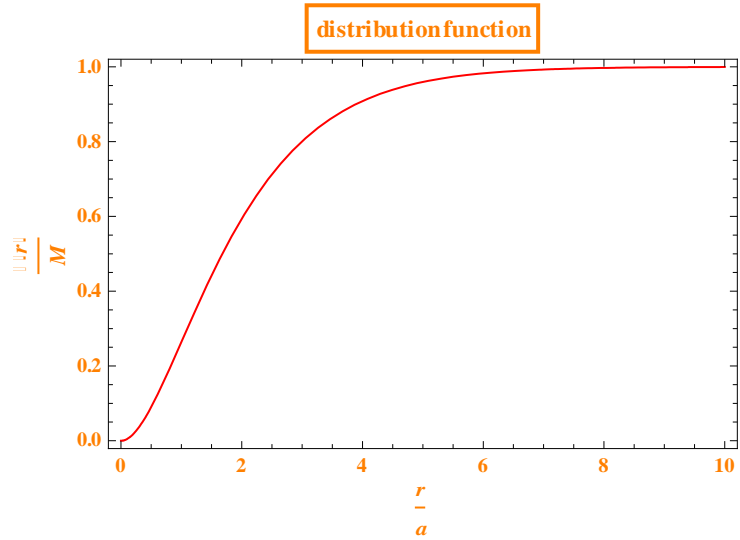

Figure2: The relation between $\frac{\mu(r)}{M}$ and $\frac{r}{a}$.

We note that at ($\frac{r}{a} \to \infty$) the distribution function tends the mass of black hole ($\frac{\mu(r)}{M} \to 1$) except at r = 0 .

Using the metric of the non-rotating BTZ black hole which is given by

$$ds^2 = -f(r)dt^2 + f^{-1}(r)dr^2 + r^2 d\varphi^2 \quad (3)$$

where

$$f(r) = -8\mu(r) + \frac{r^2}{L^2} \quad (4)$$

$L^2$ is the square of radius of curvature which related to the pressure by $L^2 = -\frac{2}{\Lambda} = \frac{1}{8\pi P}$ and $\Lambda$ is the cosmological constant; this metric reduce to the metric of ordinary BTZ black hole as r ≫ a and on other hand it goes to the anti de sitter space as r approach to zero:

$$ds^2 = \left(\frac{4Mr^2}{a^2} - \frac{r^2}{L^2}\right)dt^2 - \left(\frac{4Mr^2}{a^2} - \frac{r^2}{L^2}\right)^{-1} dr^2 + r^2 d\varphi^2 \quad (5)$$

with the new cosmological constant $\bar{\Lambda} = -\frac{4M}{a^2} + \frac{1}{L^2}$.

Putting f(r) = 0 in equation (4), then we get the mass in the form

$$M = \frac{r_H^2}{8L^2}\left[1 - e^{\frac{-r_H}{a}}\left(1 + \frac{r_H}{a}\right)\right]^{-1} \quad (6)$$

The extremal radius not depends on the vacuum pressure in non-rotating BTZ, which can be determined by $(\partial M/\partial r_H)|_{r_H=r_o} = 0$, where

$$\frac{\partial M}{\partial r_H}\bigg|_{r_H=r_o} = \frac{r_H\left[1 - e^{\frac{-r_H}{a}}\left(1 + \frac{r_H}{a} + \frac{1}{2}\left(\frac{r_H}{a}\right)^2\right)\right]}{4L^2\left[1 - e^{\frac{-r_H}{a}}\left(1 + \frac{r_H}{a}\right)\right]^2} \quad (7)$$

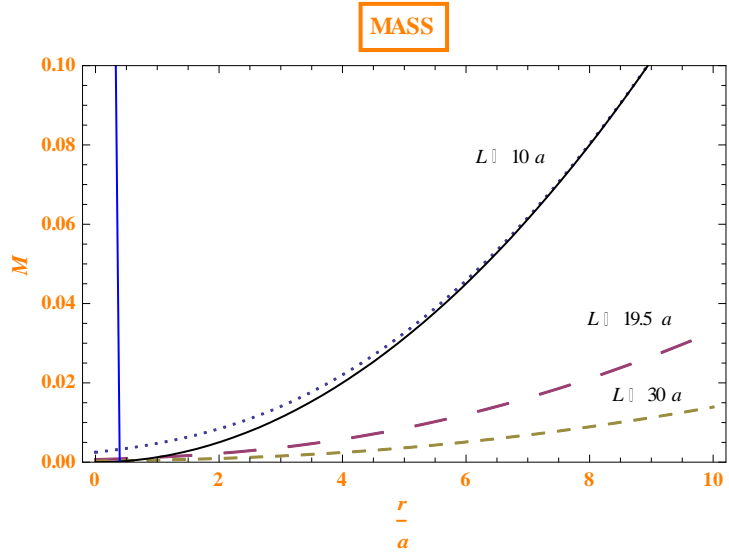

Figure 3: The relation of $M$ with respect to $r_H$; the dotted, long dashed and dashed curves correspond to the cases of $\frac{L}{a} = 10, 19.5$ and $30$ respectively, the solid line corresponds to the mass of ordinary non-rotating BTZ black hole with $M = \frac{r_H^2}{8L^2}$.

We observed from figure 3 that there is a lower bound $M_o$ for the non-rotating BTZ black hole mass, which is regarded as the mass of the extreme black hole. For the case of $M > M_0$ there exist one event horizon but for $M = M_o$, the event horizon coalesce into $r_H = r_o = 0.002a$ as the extremal radius, that is, the minimal length since no black holes are smaller than the extreme black hole.

**II B. The Equation of State and Entropy**

The Hawking temperature is given by

$$T = \frac{\kappa}{2\pi} = \frac{f'(r)}{4\pi}\bigg|_{r=r_+} \quad (8)$$

where κ is the surface gravity and $r_+$ is the outer horizon radius of the BTZ black hole.

Substituting from equations equation (2) and (4) in equations (8), we get

$$T = \frac{r_+}{2\pi L^2} - \frac{r_+^3}{4\pi L^2 a^2}\left[e^{\frac{r_+}{a}} - \left(1 + \frac{r_+}{a}\right)\right]^{-1} \quad (9)$$

From figure 4; one common property that the temperature of the extreme black hole vanishes proved from this curves, which we can show this directly by substituting (7) into (9); which proved that the evaporation of BTZ black holes behaves well; so due to the appearance of the minimal length it has no divergency [7]. Also this curves give different characteristics when the vacuum pressure increasing. In fact, (9) is the equation of state and Figure 4 is the diagram of isobar in the temperature-volume plane. One can rewrite (9) in a familiar way by using (4),

$$P = \frac{T}{r_+}\left\{4 - \frac{2r_+^2}{a^2}\left[e^{\frac{r_+}{a}} - \left(1 + \frac{r_+}{a}\right)\right]^{-1}\right\}^{-1} (10)$$

Which is plotted in figure 5 which eqn.(10) denoted the pressure of the non-rotating BTZ black hole as the diagram of isotherm in the pressure volume plane as shown in Figure 5. The diagram reveals the similarity to that of the van der Waals fluid, it is divergence at $\frac{r_o}{a} = 0.002$ which is the extremal radius so for $r_+ > r_o$ our calculation exist and for $r_+ < r_o$ not exist which is consistence with the temperature.

|Using (6), we get the thermodynamics volume at aconstant temperature in the form

$$V = \frac{\partial M}{\partial P} = \pi r_+^2 \left[1 - e^{\frac{-r_+}{a}}\left(1 + \frac{r_+}{a}\right)\right]^{-1} (11)$$

At $a \to 0$, we get $V = \pi r_+^2$, which is the thermodynamics volume of the ordinary BTZ black hole.

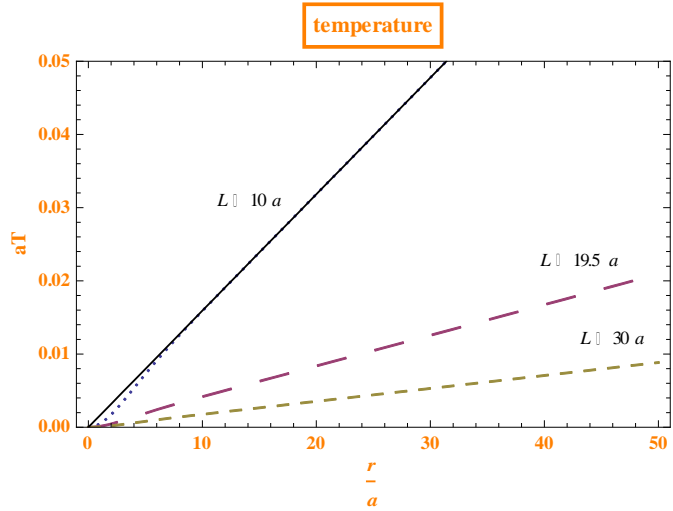

Figure 4. The Hawking temperature $T$ with respect to $r_+$ under different vacuum pressure; the dotted line, the long dashed line and the dashed line correspond to the cases of L/$a$ = 10, 19.5 and 30 respectively, the solid line is the ordinary BTZ.

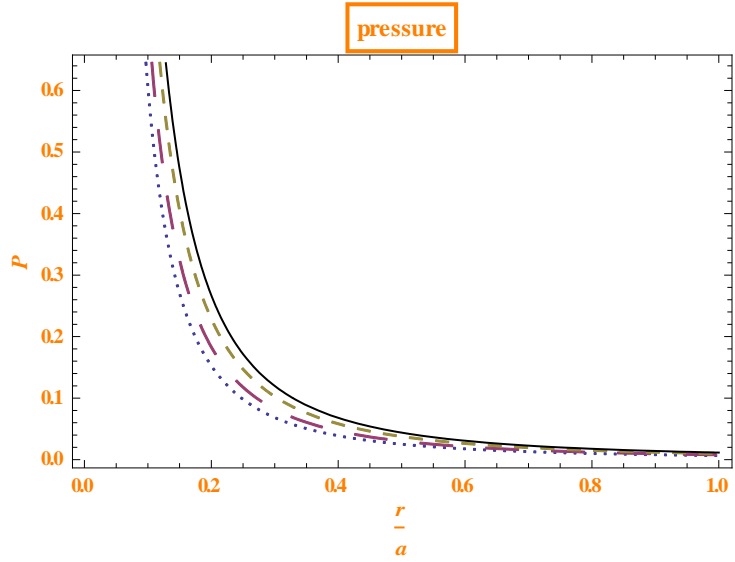

Figure 5. The equation of state; the dashed line corresponds to the cases of T=0.012, T=0.0096 (long dashed) from top to bottom respectively, the dotted line corresponds to T= 0.008, the solid line represents to the temperature T= 0.014.

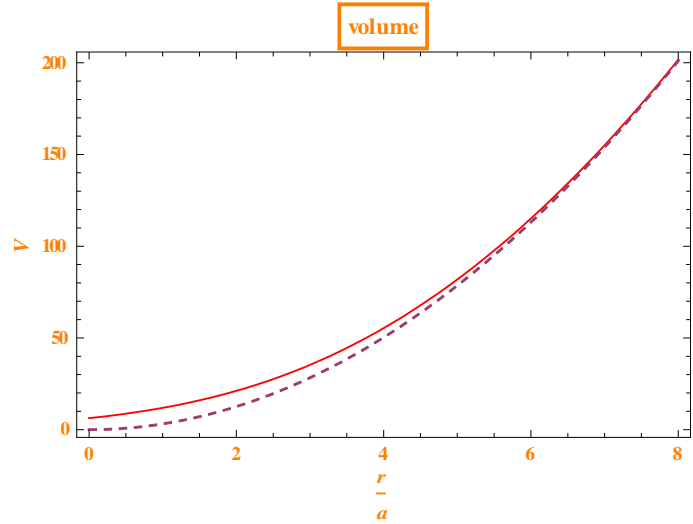

Figure 6. The thermodynamics volume with the radial coordinate; the solid line corresponding to equation (11) and the dashed line corresponding to the thermodynamics volume of the ordinary BTZ.

The mass of black hole is give by

$$dM = TdS + VdP \qquad (12)$$

Using equation (7) and (9), the entropy at fixed pressure is given by

$$S = \int_{r_o}^{r_+} \frac{dM}{T} = \int_{r_o}^{r_+} \frac{\pi dr}{2[1-e^{\frac{-r}{a}}(1+\frac{r}{a})]} \qquad (13)$$

if $a \to 0$ we have $s = \frac{1}{2}\pi r_+$.

By expanding (13) near $r \approx 0$ we get

$$S = \int_{r_o}^{r_+} \frac{\pi}{2}\left[1 + e^{-\frac{r}{a}}\left(1+\frac{r}{a}\right) - \frac{1}{2}e^{-\frac{2r}{a}}\left(1+\frac{r}{a}\right)^2 + o\left(e^{-\frac{3r}{a}}\right)\right]dr$$

Integration by parts, we get

$$S = \frac{\pi}{2}[(r_+ - r_o) - (r_+ + 2a)e^{-\frac{r_+}{a}} + \frac{1}{4}\left(3r_+ + \frac{5a}{2} + \frac{r_+^2}{a}\right)e^{-\frac{2r_+}{a}}$$
$$+ (r_o + 2a)e^{\frac{-r_o}{a}} - \frac{1}{4}\left(3r_o + \frac{5a}{2} + \frac{r_o^2}{a}\right)e^{-\frac{2r_o}{a}} + \cdots] \quad (14)$$

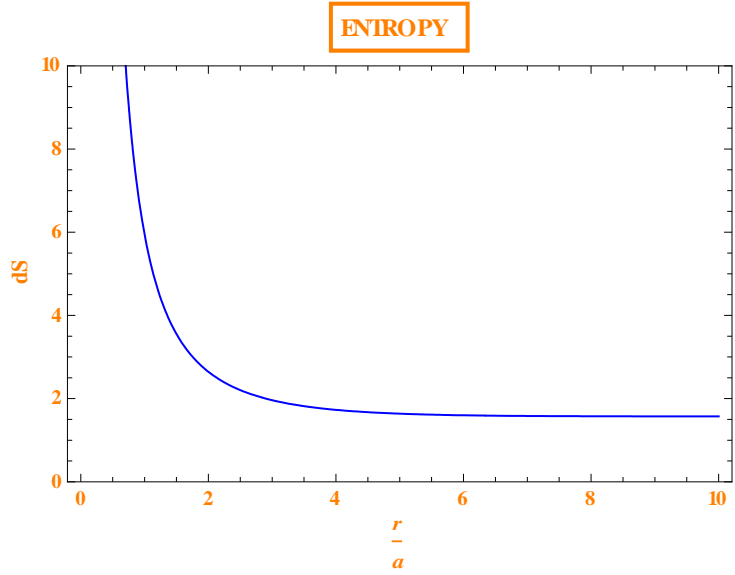

Figure 7. The entropy and $r_+$

As shown in Figure 7, we note that the entropy has deviation from the area law for the near-extreme configuration, but when the horizon radius increasing we get the deviation decreasing and finally tends to constant.

## II C. The Heat Capacity (Phase Transition)

The heat capacity at constant pressure is given by

$$C_P = \frac{\partial M}{\partial T} = \left(\frac{\partial M}{\partial r_+}\right)\left(\frac{\partial T}{\partial r_+}\right)^{-1} \quad (15)$$

Using equations (7) and (9), we get

$$\frac{\partial T}{\partial r_+} = \frac{1}{2\pi L^2} - \frac{r_+^2}{4\pi L^2 a^2}\left[e^{\frac{r_+}{a}}\left(3 - \frac{r_+}{a}\right) - 2\frac{r_+}{a} - 3\right]\left[e^{\frac{r_+}{a}} - \left(1 + \frac{r_+}{a}\right)\right]^{-2} (16)$$

$$C_P = \frac{\pi a^2 r_+ e^{\frac{r_+}{a}}[e^{\frac{r_+}{a}} - \left(1 + \frac{r_+}{a} + \frac{r_+^2}{2a^2}\right)]}{2a^2\left[e^{\frac{r_+}{a}} - \left(1 + \frac{r_+}{a}\right)\right]^2 - r_+^2\left[e^{\frac{r_+}{a}}\left(3 - \frac{r_+}{a}\right) - 2\frac{r_+}{a} - 3\right]} \quad (17)$$

when $a \to 0$ we have $C_P = \frac{\pi r_+}{2}$. We note that it diverges at the extremal points of temperature where it's sign changes from $C_P > 0$ to $C_P < 0$, or vice versa, and it is compared with the ordinary BTZ black hole. For the regular black hole there is no phase transition and also there is no phase transition in the ordinary BTZ black hole, see Figure 8.

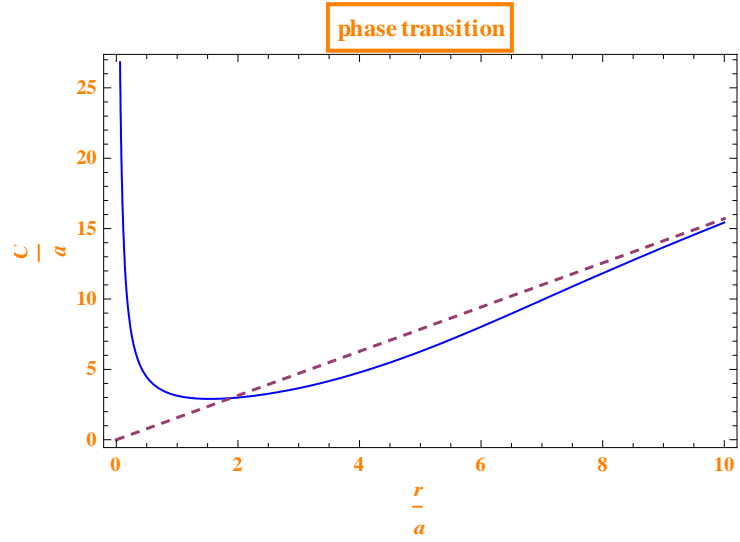

Figure 8. The heat capacity $C_P$ with respect to $r_+$; the solid line correspond to equation (17), and the dashed line corresponding to the ordinary BTZ black hole $C_P = \frac{\pi r_+}{2}$.

## III. The Collapsed Shell Model

The collapsed shell is an important model, which is usually dealt with as a massive membrane without thickness. However, the shell is supposed to be described with a smeared distribution [15], but in two dimensions if the minimal length is considered.

### III A. Mass Density Based on a Collapsed Shell:

The mass density is given by

$$\rho = \frac{100 M l_0^2 r^2}{2\pi(l_0^2 + 5r^2)^3} \qquad (18)$$

where $l_o$ is a quantity with minimal dimension which we can regard it as a minimal length and M is the total mass of BTZ black hole

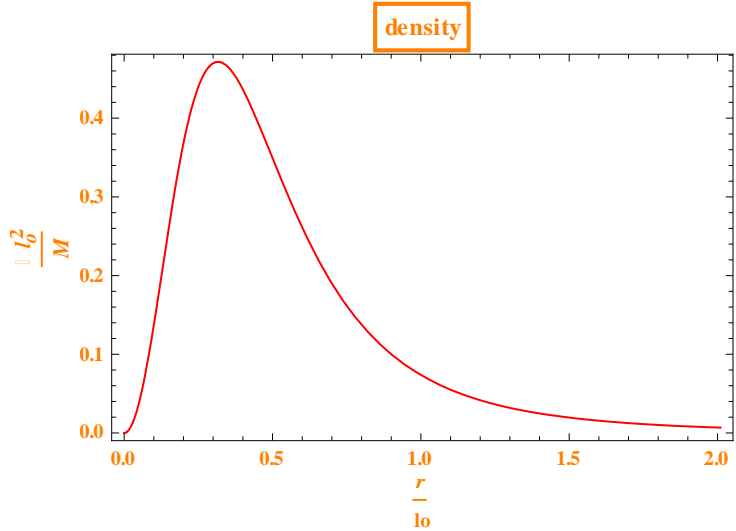

Figure 9. The mass density with the radial coordinate

The distribution function of this model is

$$\mu(r) = \frac{25Mr^4}{(l_o^2+5r^2)^2} \quad (19)$$

which like (2) we find that the distribution function tends to the mass of BTZ black hole as $l_o \to 0$ we get $\mu/M \to 1$.

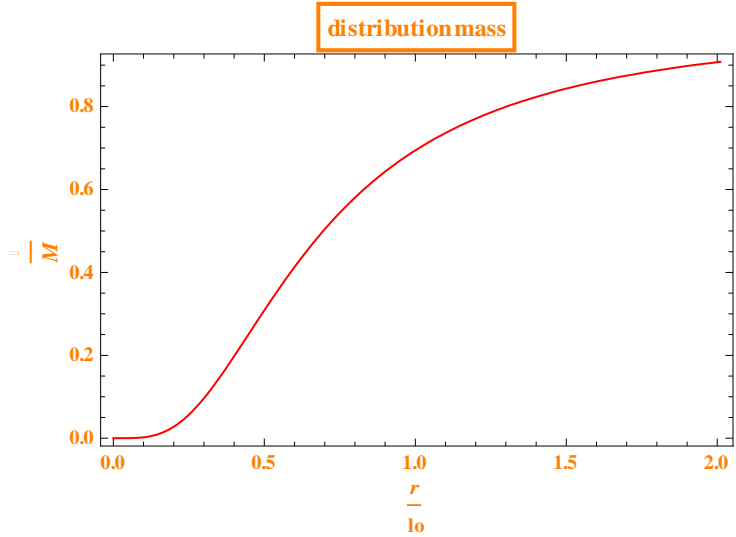

Figure 10. The distribution function with the radial coordinate.

Putting $f(r) = 0$ in equation (4), we get the mass of the BTZ black hole in terms of $r_H$ is given by

$$M = \frac{(l_o^2+5r_H^2)^2}{200L^2 r_H^2} \quad (20)$$

We noted that there is a lower bound $M_o$ for the non-rotating BTZ black hole mass, which is regarded as the mass of the extreme black hole. For the case of $M > M_0$, we have two event horizon and the case of $M = M_o$, the two horizons coalesce into one, $r_H = r_o = 0.45\, l_o$, as the extremal radius, that is, the minimal length since no black holes are smaller than the extreme black hole. The extremal radius does not depend on the vacuum pressure in non-rotating BTZ, which can be determined by $(\partial M/\partial r_H)|_{r_H=r_o} = 0$, where

$$(\partial M/\partial r_H)|_{r_H=r_o} = 0 = \frac{(l_o^2+5r_H^2)(-l_o^2+5r_H^2)}{100L^2 r_H^3} \quad (21)$$

So the extremal radius is

$$r_o = \frac{l_o}{\sqrt{5}} \quad (22)$$

which is constant.

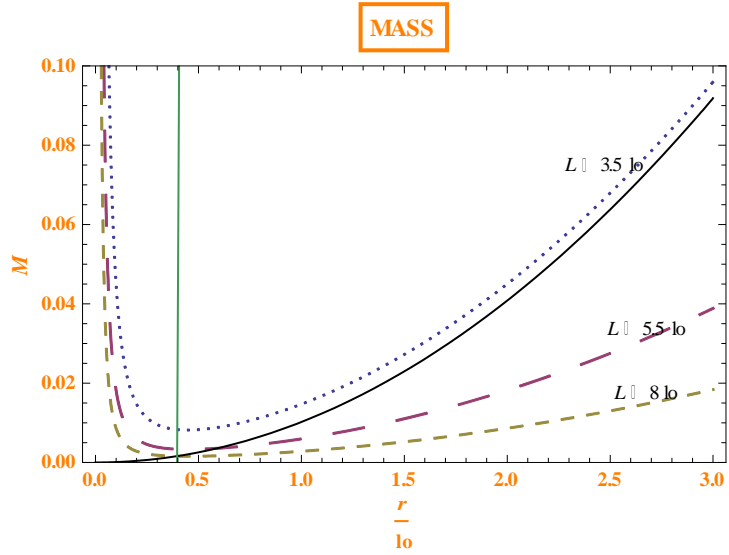

Figure 11. The relation between M and $\frac{r_H}{l_o}$; the dotted line, long dashed line and dashed line corresponding to $\frac{L}{l_o} = 3.5, 5.5$ and $8$ respectively, and the vertical line lies at the extremal radius, the solid line corresponding to ordinary BTZ.

Using equation (18) the mass density, we get the Hawking temperature in the form

$$T = \frac{r_+(-l_0^2 + 5r_+^2)}{2\pi L^2(l_0^2 + 5r_+^2)} \quad (23)$$

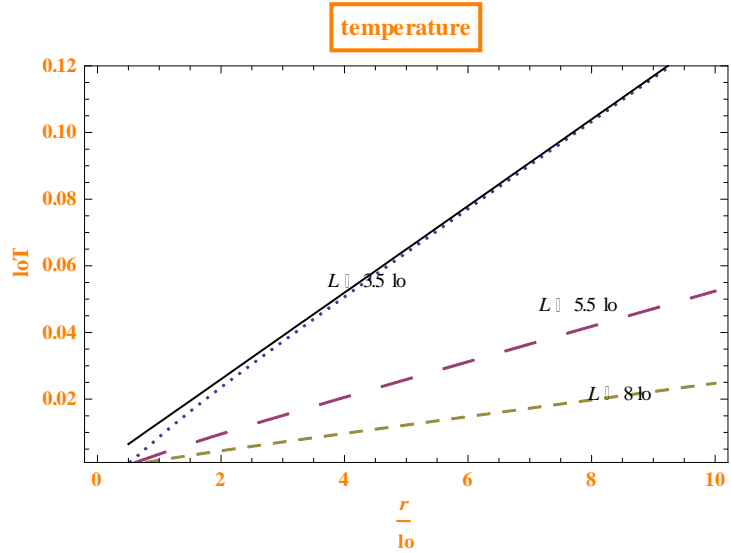

Figure 12: The Hawking temperature $T$ with respect to $r_+$ under different vacuum pressure; the dotted line, the long dashed line and dashed line correspond to the cases of $L/l_o$ = 3.5, 5.5 and 8 respectively, the solid line corresponding to ordinary BTZ.

We note that the temperature vanishes at the extremal radius

The thermodynamics volume for the BTZ black hole of the mass density for collapsed shell model is

$$V = \frac{\partial M}{\partial P} = \frac{\pi(l_o^2+5r_+^2)^2}{25r_+^2} \quad (24)$$

at $l_o \to 0$ we get the thermodynamics volume of ordinary BTZ

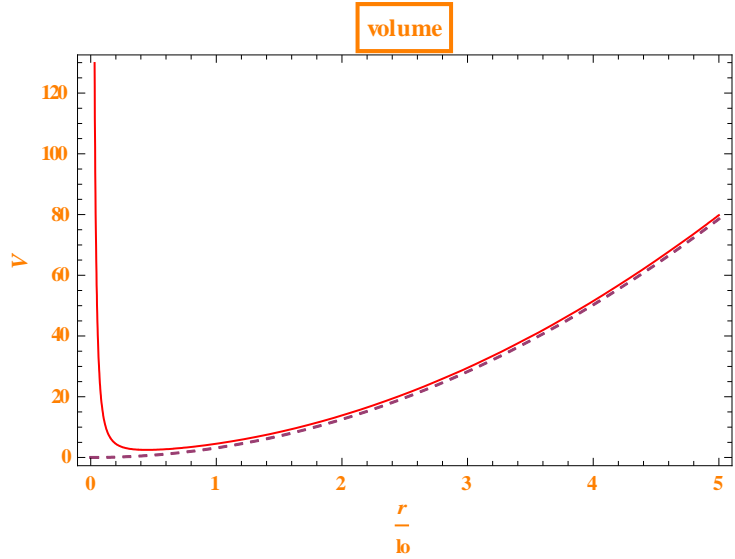

Figure 13. The thermodynamics volume with $r_+$; the dashed line is the thermodynamics volume of ordinary BTZ and the solid line corresponding to equation (24)

From figure 13, we note that the volume divergence at the extremal radius.

Using equations (12) and (23) we get the entropy in the form

$$S = \frac{\pi}{2}\left[(r_+ - r_o) - \frac{2l_o^2}{5}\left(\frac{1}{r_+} - \frac{1}{r_o}\right) - \frac{l_o^4}{75}\left(\frac{1}{r_+^3} - \frac{1}{r_o^3}\right)\right] (25)$$

Also the first term is the horizon area of BTZ black hole satisfies the area law of the ordinary BTZ black hole thermodynamics, and the other terms are quantum corrections which are induced by the minimal length.

The heat capacity at constant pressure is

$$C_P = \frac{\pi(l_o^2+5r_+^2)^3(-l_o^2+5r_+^2)}{50r_+^3(25r_+^4+20r_+^2l_o^2-l_o^4)} \quad (26)$$

which also does not depend on the pressure and when $l_o \to 0$ we get the heat capacity for ordinary BTZ black hole $C_P = \frac{\pi r_+}{2}$. For the regular black hole there is one first-order phase transition near the extremal point.

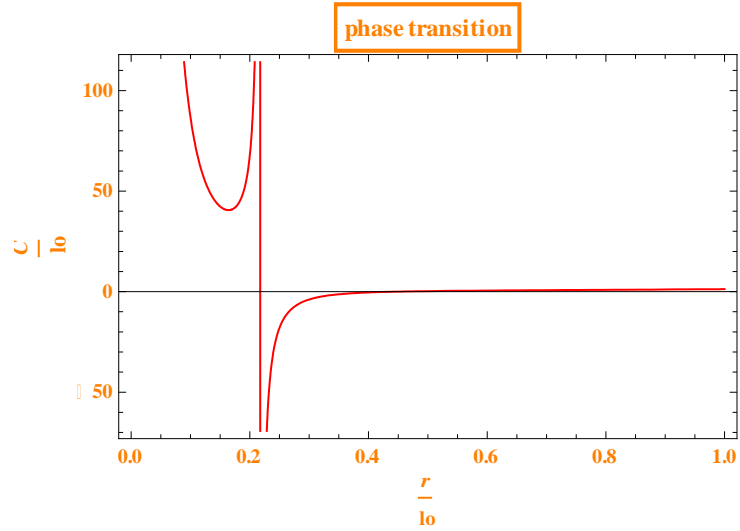

Figure 14. The heat capacity $C_P$ with respect to $r_+$.

## IV. The Thermodynamics cycle and heat engine of BTZ

Finally in ordinary BTZ black hole we can calculate some of thermodynamics quantity by using (temperature, volume, entropy and pressure) such as the heat capacity at constant pressure and the work. We first start our calculation with the equation of state and then we define the BTZ black hole as a heat engine which define as a closed path in $P - V$ plane which receives $Q_H$ and denote $Q_C$, from the first law of thermodynamics the total work W for any heat engine is denoted by $W = Q_H - Q_C$. We know that the efficiency is define by $\eta = \frac{W}{Q_H} = 1 - \frac{Q_C}{Q_H}$.

We know that we can relate two systems with each other by using different methods. The first method is isochoric path like classical Stirling cycle and other cycle which based on adiabatic path like Carnot cycle(we know that its heat engine is full reversible since the total entropy vanish). Consider a classical cyclic contains pair of isothermal at different temperature $T_H$ and $T_C$ ($T_H > T_C$).

We know that the Carnot efficiency is maximum efficiency which denoted by

$$\eta_C = \frac{W}{T_H} = 1 - \frac{T_C}{T_H} \quad (27)$$

For any higher efficiency it would violate the second Law. The form of the path for the definition of the cycle is very important because we want to know that how we can reach to this efficiency in heat engine of black holes to preserve the second law.

As we know, for any BTZ black holes the thermodynamics volume V and entropy S are dependent. In this case Carnot and Stirling coincide to each other due to the diabetes and isochors are the same. So, we can calculate the efficiency of cycle easily.

Since the event horizon is a circle and the entropy is quarter of it's circumference

$$S = \frac{2\pi r_+}{4} = \frac{\pi r_+}{2} \quad (28)$$

Thermodynamics volume is

$$V = \pi r_+^2 \quad (29)$$

From the second law of thermodynamics we have the upper and lower isotherm are given in the following forms respectively

$$Q_H = T_H \Delta S_{1 \to 2} = T_H \left(\frac{\pi}{4}\right)^{\frac{1}{2}} (V_2^{\frac{1}{2}} - V_1^{\frac{1}{2}})$$

$$Q_C = T_C \Delta S_{3 \to 4} = T_C \left(\frac{\pi}{4}\right)^{\frac{1}{2}} \left(V_3^{\frac{1}{2}} - V_4^{\frac{1}{2}}\right) \quad (30)$$

But since $V_1 = V_4$ and $V_2 = V_3$ then we get

$$\eta = \frac{W}{T_H} = 1 - \frac{T_C}{T_H} \quad (31)$$

which is the Carnot engine

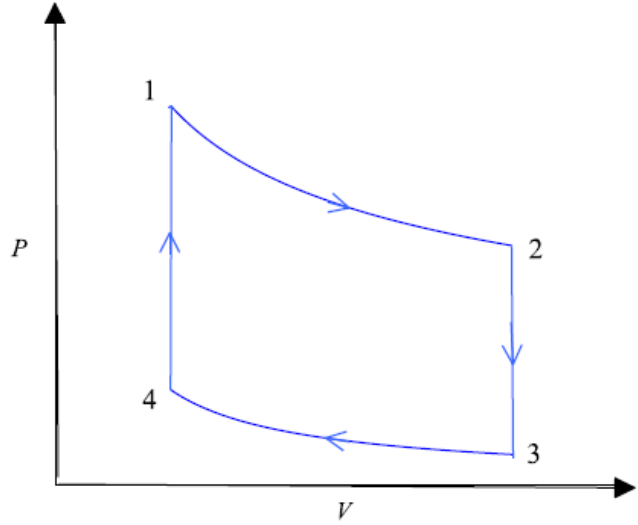

Figure 15. The Carnot heat engine

So, we take advantage from above thermodynamics cycles and heat engines information and study the efficiency for varies BTZ black hole.

## V. The efficiency, thermodynamics variables of static uncharged BTZ black holes and heat engine

Using equation (3) in ordinary state of BTZ black hole (i.e. $a \to 0$) we have

$$ds^2 = -f(r)dt^2 + f^{-1}(r)dr^2 + r^2 d\varphi^2 \quad (32)$$

where

$$f(r) = -8M(r) + \frac{r^2}{L^2} \quad (33)$$

Now we want the obtain the equation of state, so we started with the temperature

$$T = \frac{\kappa}{2\pi} = \frac{f'(r)}{4\pi}\big|_{r=r_+} = \frac{r_+}{2\pi L^2} = \frac{S}{\pi^2 L^2} \quad (34)$$

the thermodynamics volume in terms of the entropy is

$$V = \frac{4S^2}{\pi} \quad (35)$$

So, the equation of state is

$$T = 4P\left(\frac{V}{\pi}\right)^{\frac{1}{2}} \quad (36)$$

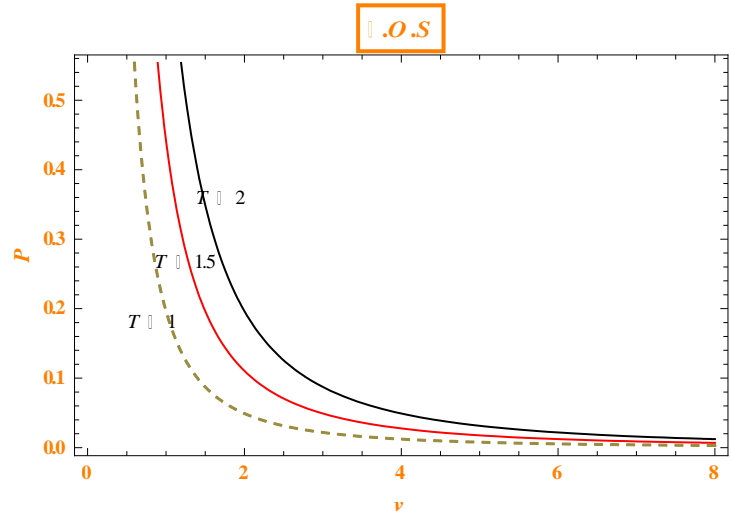

Figure 16. The equation of state (36) for varies values of temperature

from figure 16 we have two different solutions of black hole; the first is the small BTZ black hole (SBH) in the left for large P. But in the right where law p, we have the large BTZ black hole (LBH) so from this we have SBH tends to LBH. If the large reduce to small then we can define classical cycle for this black hole. The small black hole absorbs heat $Q_H$ along isothermal expansion and exist at high pressure.

The heat capacity is an important measurable physical quantity; it determines the amount of requisite heat to change the temperature of an object by a given amount. There are two different heat capacities for a system, heat capacity at constant pressure and heat capacity at constant volume. Heat capacity can be calculated by the standard thermodynamics relations, which is given by,

$$C_V = T\frac{\partial S}{\partial T}\Big|_V \quad \text{and} \quad C_P = T\frac{\partial S}{\partial T}\Big|_P \quad (37)$$

From equation (35) at constant volume we note that the entropy is constant so $C_V = 0$. But by using equation (34) we get

$$C_P = \frac{\pi T}{8P} \quad (38)$$

Actually, an explicit expression $C_P$ lead us to have a new engine which include two isobars and two isochores/adiabatas as figure(17).

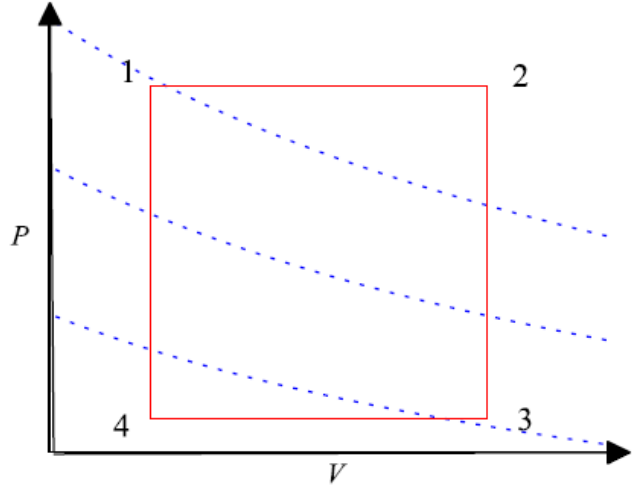

Figure 17: our other cycle

Now the work done in this cycle is

$$W_{total} = \oint PdV = W_{1\to 2} + W_{3\to 4}$$
$$= P_1(V_2 - V_1) + P_4(V_3 - V_4)$$
$$= \frac{4}{\pi}(S_2^2 - S_1^2)(P_1 - P_4) \quad (39)$$

From the heat capacity (upper isobar) we obtain

$$Q_H = \int_{T_1}^{T_2} C_P(P_1, T)\, dT \quad (40)$$

So, by using equation (38) we get

$$Q_H = \frac{\pi}{16P_1}(T_2^2 - T_1^2) = \frac{4P_1}{\pi}(S_2^2 - S_1^2) \quad (41)$$

From equations (39) and (41) we get the efficiency of the BTZ in the form

$$\eta = \frac{W}{Q_H} = \left(1 - \frac{P_4}{P_1}\right) = 1 - \frac{T_C}{T_H} = \eta_C \quad (42)$$

The above equation is the efficiency of the static uncharged BTZ black hole by using the Carnot cycle, which is consistence with the second law of thermodynamics.

## VI. The efficiency, thermodynamics variables of charged and rotating BTZ black holes and heat engine

Now we want to study the thermodynamics variables and the efficiency for three type of BTZ black hole the first for charged and non-rotating BTZ, the second for rotating and uncharged, finally in general case respectively. We started our calculation by the line element which is given by the following metric [16, 17] in general

$$(43)\, ds^2 = -f(r)dt^2 + f^{-1}(r)dr^2 + r^2\left[d\varphi - \frac{4J}{r^2}dt\right]^2$$

where

$$(44) f(r) = \left[-8M + \frac{r^2}{L^2} + \frac{16J^2}{r^2} - 32\pi q^2 \ln r\right]$$

where J is angular momentum and q is the charge.

## VI A. Thermodynamics variables and efficiency of charged non-rotating BTZ black hole: $(Q \neq 0 \text{ and } J = 0)$

For non-rotating black hole by setting $J = 0$ in eqn.(43) then becomes

$$(45) ds^2 = -f(r)dt^2 + f^{-1}(r)dr^2 + r^2 d\varphi^2$$

where

$$(46) f(r) = \left[-8M + \frac{r^2}{L^2} - 32\pi q^2 \ln r\right]$$

We started our calculation by obtain Hawking temperature

$$T = \frac{\kappa}{2\pi} = \frac{f'(r)}{4\pi}\bigg|_{r=r_+}$$

$$(47) T = \frac{r_+}{2\pi L^2} - \frac{8q^2}{r_+}$$

But by using (28) we obtain the equation of state in terms of $(S, P)$ as

$$(48) T = \frac{8PS}{\pi} - \frac{4\pi q^2}{S}$$

And the equation of state in terms of $(P, V)$ substituting from (29) into (47) we get

$$(49) T = \sqrt{\frac{V}{\pi}} 4P - 32 q^2 \sqrt{\frac{\pi^3}{V}}$$

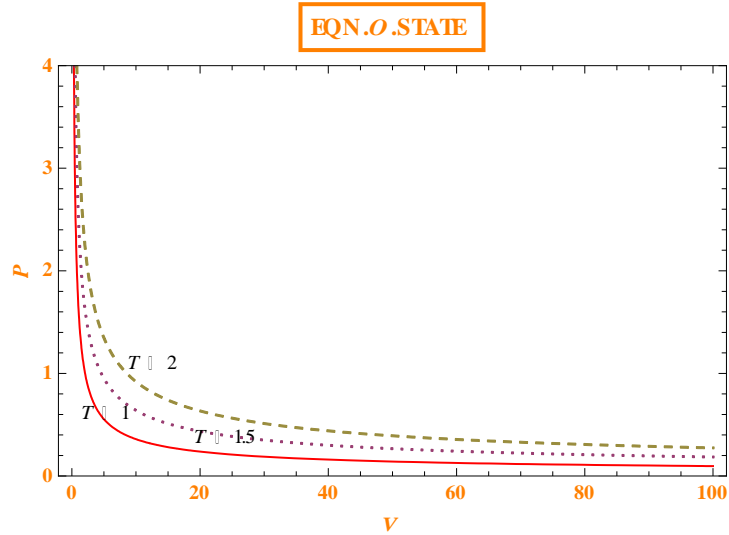

Figure 18. The pressure with volume at fixed q = 0.1 equation (49)

It is obvious that the curve reduce from small black hole to large so we can define the heat engine for charged BTZ black hole. So, this property leads us to assume that the BTZ black hole as Carnot cycle and then we obtain the efficiency.

Now the heat capacity at constant volume also equal zero due to the same reason above. But we want to calculate the heat capacity at constant pressure by using (46) we get

$$C_P = T\frac{\partial S}{\partial T}|_P = T\left(\frac{8P}{\pi} + \frac{4\pi q^2}{S^2}\right)^{-1} \quad (50)$$

$$C_P = \frac{\pi T}{8P}\left[1 - \frac{\pi^2 q^2}{2PS^2} + \frac{\pi^4 q^4}{4P^2 S^4} + O\left(\frac{q^6}{P^3 S^6}\right)\right] \quad (51)$$

We take the high entropy limit to deduce explicit form for the efficiency. But to obtain the heat of engine we must represent the heat capacity in terms of the temperature.

So, by using the relation between T and S for high entropy we get

$$C_P = \left[\frac{\pi T}{8P} - \frac{4\pi q^2}{T} + \frac{128\pi P^2 q^4}{T^3} + O\left(\frac{q^6}{P^3 S^6}\right)\right] \quad (52)$$

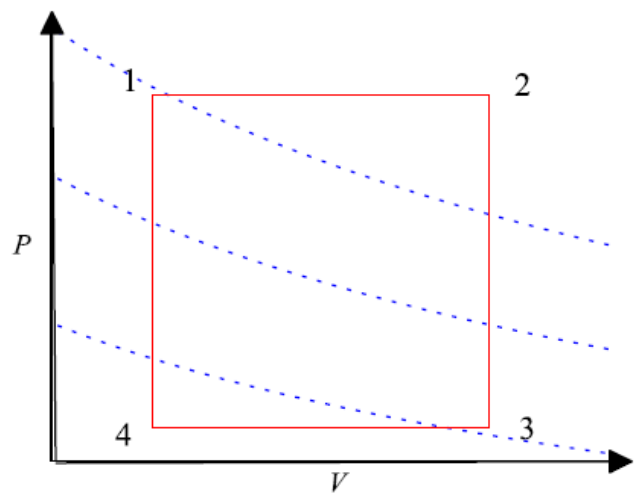

Figure 19. Our other engine

By using the definition of heat (the upper isobar)

$$Q_H = \int_{T_1}^{T_2} C_P(P_1, T)\, dT \quad (Eqn.40)$$

So, by using (52) we get

$$Q_H = \left[\frac{\pi}{16P_1}(T_2^2 - T_1^2) - 4\pi q^2 \ln\left(\frac{T_2}{T_1}\right) - \frac{64P^2(\pi q^2 - 3J^2)}{T_2^2 - T_1^2} + \cdots\right] (53)$$

By using the relation between the entropy and temperature

$$Q_H = \left[\frac{4P_1}{\pi}(S_2^2 - S_1^2) - 4\pi q^2 \ln\left(\frac{S_2}{S_1}\right) - \frac{\pi^3 q^2}{(S_2^2 - S_1^2)} + \cdots\right] \quad (54)$$

Now by using (39) and (54) we get the efficiency for high entropy is

$$\eta = \frac{W}{Q_H} = \left(1 - \frac{P_4}{P_1}\right)\left[\frac{(S_2^2 - S_1^2)}{\frac{4P_1}{\pi}(S_2^2 - S_1^2) - \frac{\pi^2 q^2}{P_1}\ln\left(\frac{S_2}{S_1}\right) - \frac{4q^2}{4P_1(S_2^2 - S_1^2)} + \cdots}\right] \quad (55)$$

By using the equation of state we get for high pressure we get

$$\eta = \left(1 - \frac{T_C}{T_H}\right)\left[1 + \frac{\pi^2 q^2 \ln\left(\frac{S_2}{S_1}\right)}{P_1(S_2^2 - S_1^2)} + \frac{\pi^4 q^2}{4P_1(S_2^2 - S_1^2)^2} + \cdots\right] \quad (56)$$

## VI B. Thermodynamics variables and efficiency of uncharged rotating BTZ black hole: ($Q = 0$ and $J \neq 0$)

So, equation (43) becomes

$$(57)\, ds^2 = -f(r)dt^2 + f^{-1}(r)dr^2 + r^2\left[d\varphi - \frac{4J}{r^2}dt\right]^2$$

where

$$(58)\, f(r) = \left[-8M + \frac{r^2}{L^2} + \frac{16J^2}{r^2}\right]$$

We can obtain the equation of state by introducing the Hawking temperature

$$(59)\, T = \frac{1}{4\pi}\left(\frac{2r_+}{L^2} - \frac{32J^2}{r_+^3}\right)$$

Then we have the temperature as a function of $(S, P)$ is

$$(60)\, T = \frac{8PS}{\pi} - \frac{\pi^2 J^2}{S^3}$$

And the equation of state in terms of $(P, V)$ substituting from (29) into (60) we get

$$(61)\, P = \frac{1}{4}\sqrt{\frac{\pi}{V}}T + \frac{4\pi J^2}{V^2}$$

See figure20.

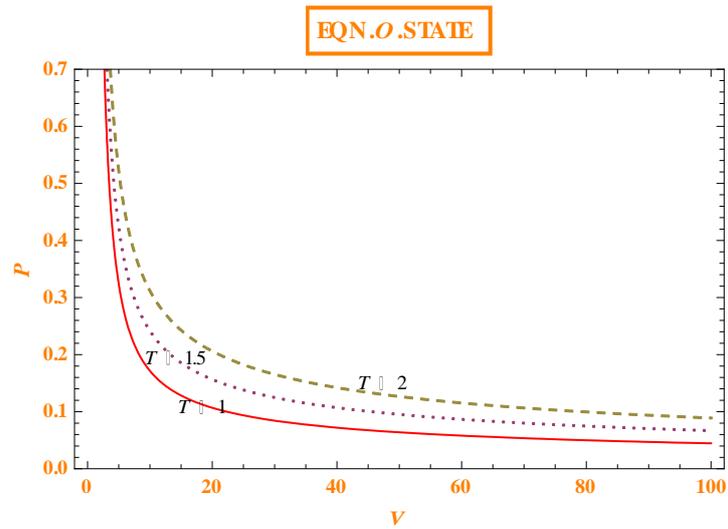

Figure 20 : P-V equation of state (61) for fixed $J = 0.5$

From this figure we note that this also define heat engine. So,

The heat capacity at constant pressure for high entropy by using (60) we get

$$C_P = T\frac{\partial S}{\partial T}|_P = T\left(\frac{8P}{\pi} + \frac{3\pi^2 J^2}{S^4}\right)^{-1} \quad (62)$$

$$C_P = \frac{\pi T}{8P}\left[1 - \frac{3\pi^3 J^2}{8PS^4} + O\left(\frac{J^4}{P^2 S^8}\right)\right] (63)$$

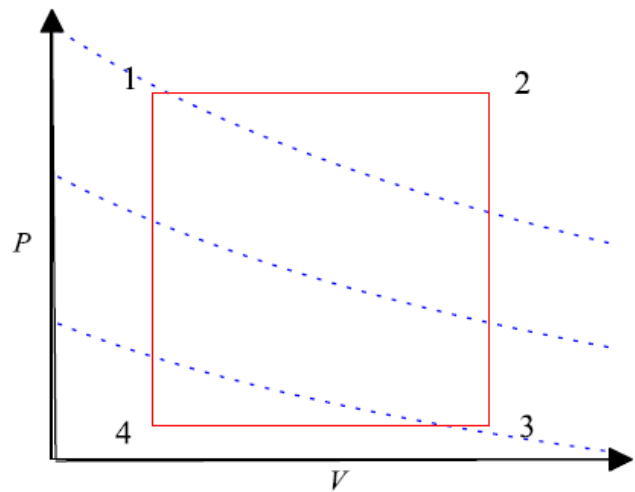

Figure 21 :our other engine

But to obtain the heat of the engine we must represent the heat capacity in terms of the temperature

So, by using the relation between T and S for high entropy equation (61) we get

$$C_P = \left[\frac{\pi T}{8P} - \frac{192 P^2 J^2}{T^3} + O\left(\frac{P^5 J^4}{T^7}\right)\right] (64)$$

We know that

Using (64) we get

$$Q_H = \left[\frac{\pi}{16P_1}(T_2^2 - T_1^2) + \frac{96P^2J^2}{T_2^2-T_1^2} + \cdots\right] \quad (65)$$

By using the relation between the entropy and temperature

$$Q_H = \left[\frac{4P_1}{\pi}(S_2^2 - S_1^2) + \frac{3\pi^2J^2}{2(S_2^2-S_1^2)} + \cdots\right] \quad (66)$$

Now by using (39) and (66) we get the efficiency for high entropy and high pressure is

$$\eta = \left(1 - \frac{T_C}{T_H}\right)\left[1 + \frac{3\pi^3J^2}{8P_1(S_2^2-S_1^2)^2} + \cdots\right] \quad (67)$$

## VI C. Thermodynamics variables and efficiency of charged rotating BTZ black hole: $(Q \neq 0 \text{ and } J \neq 0)$

From equation (43) we get the Hawking temperature in the form

$$(68) T = \frac{1}{4\pi}\left(\frac{2r_+}{L^2} - \frac{32J^2}{r_+^3} - \frac{32\pi q^2}{r_+}\right)$$

The temperature in terms of the thermodynamics variable is

$$(69) T = \frac{8PS}{\pi} - \frac{\pi^2J^2}{S^3} - \frac{4\pi q^2}{S}$$

And the equation of state in terms of $(P, V)$ substituting from (29) into (69) we get

$$(70) P = \frac{1}{4}\sqrt{\frac{\pi}{V}}T + \frac{4\pi J^2}{V^2} + \frac{8\pi^2 q^2}{V}$$

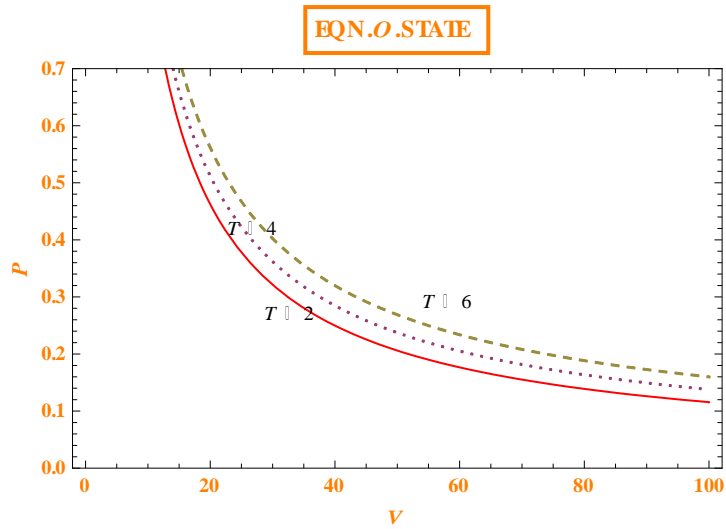

Figure 22. P-V diagram, which represent the equation of state equation (70) by varying the temperature $J = q = 0.1$

It is very important to calculate the heat capacity at constant pressure by using equation (69) we get

$$C_P = T\frac{\partial S}{\partial T}\big|_P = T\left(\frac{8P}{\pi} + \frac{3\pi^2 J^2}{S^4} + \frac{4\pi q^2}{S^2}\right)^{-1} \quad (71)$$

$$C_P = \frac{\pi T}{8P}\left[1 - \frac{3\pi^3 J^2}{8PS^4} - \frac{\pi^2 q^2}{2PS^2} + \frac{\pi^4 q^4}{4P^2 S^4} + O\left(\frac{J^2 q^2}{P^2 S^6}, \frac{J^4}{P^2 S^8}\right)\right] \quad (72)$$

But to obtain the heat of the engine, we must represent the heat capacity in terms of the temperature
So, by using the relation between T and S for high entropy we get

$$C_P = \left[\frac{\pi T}{8P} - \frac{4\pi q^2}{T} + \frac{64P^2(2\pi q^4 - 3J^2)}{T^3} + O\left(\frac{q^2 J^2 P^2}{T^5}, \frac{P^5 J^4}{T^7}\right)\right] \quad (73)$$

Using (73) we get

$$Q_H = \left[\frac{\pi}{16P_1}(T_2^2 - T_1^2) - 4\pi q^2 \ln\left(\frac{T_2}{T_1}\right) - \frac{32P^2(2\pi q^4 - 3J^2)}{T_2^2 - T_1^2} + \cdots\right] \quad (74)$$

By using the relation between the entropy and temperature

$$Q_H = \left[\frac{4P_1}{\pi}(S_2^2 - S_1^2) - 4\pi q^2 \ln\left(\frac{S_2}{S_1}\right) - \frac{\pi^2(2\pi q^4 - 3J^2)}{2(S_2^2 - S_1^2)} + \cdots\right] \quad (75)$$

Using equations (39) and (75) we get the efficiency at high pressure in the following form

$$\eta = \left(1 - \frac{T_C}{T_H}\right)\left[1 + \frac{\pi^2 q^2 \ln\left(\frac{S_2}{S_1}\right)}{P_1(S_2^2 - S_1^2)} + \frac{\pi^3(2\pi q^4 - 3J^2)}{8P_1(S_2^2 - S_1^2)^2} + \cdots\right] \quad (76)$$

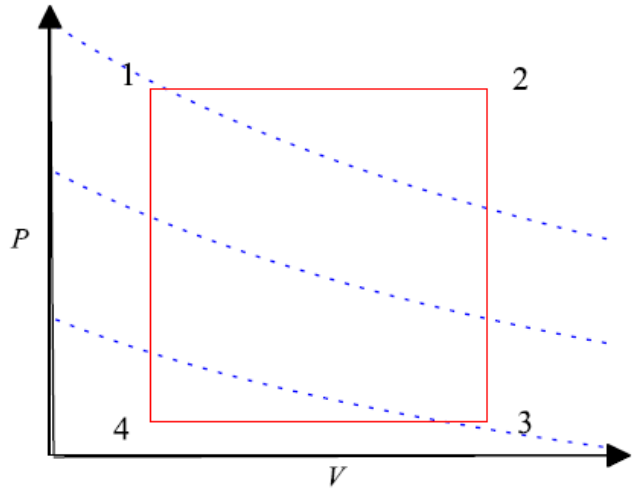

Figure 23. Our engine for this model

We note that this efficiency tends to Carnot engine for small q and J ; from figure 24, we plot the efficiency corresponding three cases of pressure ($P_1 = 0.1$, $P_1 = 0.2$, $P_1 = 0.3$). Corresponding this cases we have for $P_1 = 0.1$ we get from the range $0 < q < 3.41$ our efficiency large than Carnot efficiency which is inconsistence with the second law of thermodynamics. But for q > 3.41 we get it is consistence. At $P_1 = 0.2$ and $P_1 = 0.3$ it is inconsistence with second law in range of $0 < q < 2.63$ and $0 < q < 1.6$ respectively. But in all cases the efficiency satisfies the second law from $3.41 \leq q \leq 4.1$. For q = 4.1 our efficiency equal to Carnot efficiency where is maximum efficiency but for q > 4.1 the efficiency do not satisfy the second law of thermodynamics.

Also from figure (25) we note that at $P_1 = 0.1$ the efficiency is consistence with the second law in the range of $0 < J \leq 3.2$ and for $J > 3.2$ inconsistence due to the efficiency is large than the Carnot cycle. For $P_1 = 0.2$ our efficiency is consistence with the second law in the range of $0 < J \leq 4.37$ and otherwise inconsistence. Finally at $P_1 = 0.3$ it is consistence in the range of $0 < J \leq 5.4$ and for $J > 5.4$ inconsistent and maximum efficiency at $J = 0$ in all cases. But in figure (26) show that the angular momentum doesn't effect in the efficiency for small charge for $J \leq 1$ and denoted Carnot efficiency.

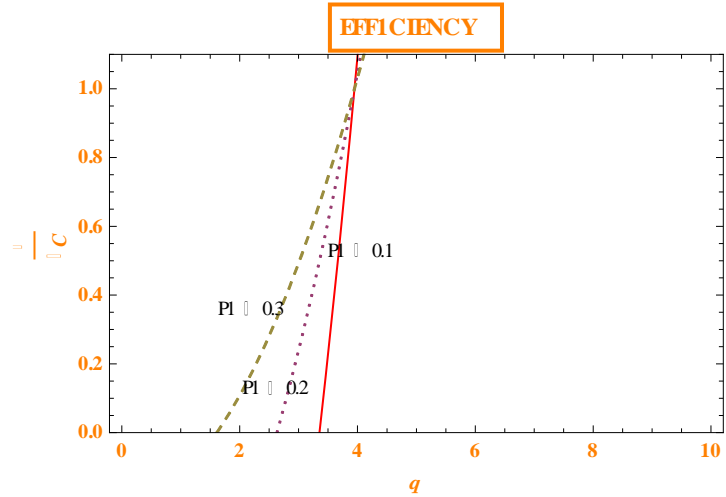

Figure 24. The ratio of efficiency with charge; $S_1 = 10$, $S_2 = 20$ and $J = 6$

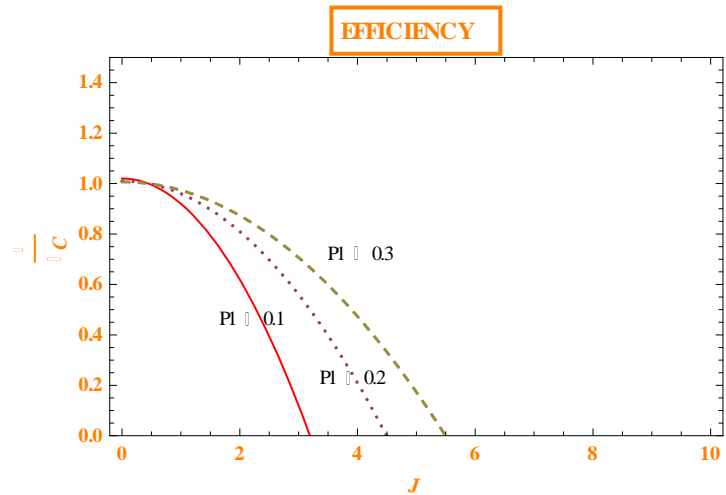

Figure 25. The relation between the efficiency and angular momentum;
$S_1 = 10$, $S_2 = 20$ and $q = 0.2$

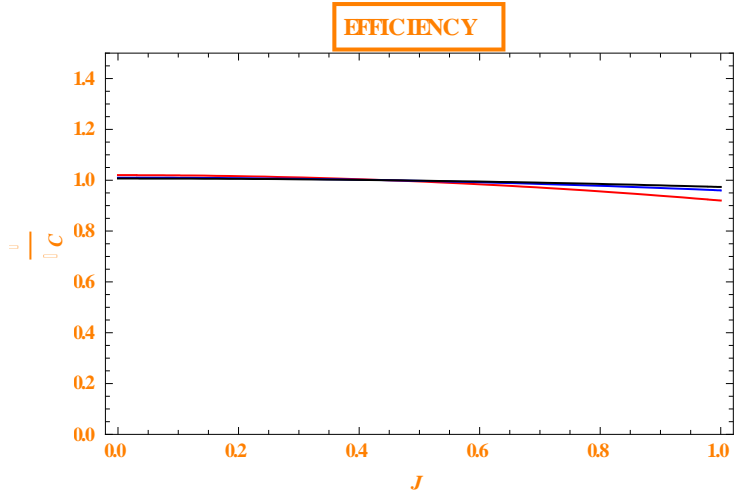

Figure 26. The relation between the efficiency and angular momentum;
$S_1 = 10$, $S_2 = 20$ and $q = 0.2$

# VII. Conclusion

In this paper we study the thermodynamics variables of the BTZ black hole with minimal length by using mass density of two models. Also we study thermodynamics properties of BTZ black hole as thermodynamics cycles and heat engine generally. We can arrange the BTZ black hole as Carnot cycle by this thermodynamics cycles and the heat engine.

The first model is the quantum hydrogen atom, which is plotted in figure (1) and in figure (2) we plotted the distribution function of this model with the radial coordinate. For this model we start our discussion by using the line element of BTZ black hole which is the solution of modified Einstein's equations corresponding to this distribution function. We can easily prove that this metric reduce to the metric of ordinary BTZ black hole as $r \gg a$ and on other hand it goes to the anti de sitter space as $a$ approach to zero with the new cosmological constant $\overline{\Lambda} = \frac{M}{2a^2} + \frac{1}{L^2}$.

In first we putting $f(r) = 0$ to obtain the mass of the BTZ black hole in terms of the radial coordinate of black hole which potted in figure (3) and then we find the extremal black hole which depends on the extremal radius which can be determined by $\left(\frac{\partial M}{\partial r_H}\right)|_{r_H = r_o} = 0$ and we have that the extremal radius is $\frac{r_o}{a} = 0.002$. Then we calculate the Hawking temperature to obtain the equation of motion which they are plotted in figures (4 and 5) respectively. And then we obtain the thermodynamics volume, plotted it in figure (6) and from the first law of thermodynamics we obtain the entropy which is very difficult to integrate it analytically so, we plot it easily in figure (7). Finally for this model we calculate the heat capacity at constant pressure due to discuss the phase transition.

In the second section we study the other models (collapsed shell model) and we plotted its mass density in figure (9) and its distribution mass in figure (10) also by setting $f(r) = 0$ we obtain the mass of black hole which corresponding to this model and plotted it in figure (11) and we noted that For the case of $M > M_0$, we have two event horizon and the case of $M = M_0$, the two horizons coalesce into one, $r_H = r_o = 0.45 l_o$ and then we obtain the temperature due to obtain the equation of state and plotted it in figure (12) and the thermodynamics volume in figure (13) but in this model we calculate the entropy exact finally for this model we obtain the heat capacity at constant pressure which plotted in figure (14) and we note that the phase transition happened at the extremal radius.

In section three we study the ordinary BTZ black hole as a heat engine by using the temperature, pressure, entropy and volume we can calculate the equation of state and the heat capacity so, we can obtain the heat of the system and a heat engine which define as a closed path in $P - V$ plane which receives $Q_H$ and denote $Q_C$ we can obtain the work which denoted by $W = Q_H - Q_C$ so we can obtain the efficiency by using $\eta = \frac{W}{Q_H} = 1 - \frac{Q_C}{Q_H}$ and we know that the Carnot efficiency is maximum efficiency which denoted by $\eta_C = 1 - \frac{T_C}{T_H}$. We start with the second law of thermodynamics which define under Carnot cycle plotted in figure (15) and we start our discussion for the static uncharged BTZ black hole, obtain the equation of motion figure (16) and we obtain the efficiency under other cycle figure (17) which we prove that the efficiency of this black hole under this cycle is Carnot efficiency.

In the final section we study the equation of state and then we obtain the upper heat to obtain the efficiency for three type of BTZ black hole charged and non-rotating plotted in figure (18), uncharged and rotating in figure (20), charged and rotating in figure (22) and we ploted in general case the equation of state in figure (18) and the efficiency in figures (24, 25 and 26) which our other engine in figures (17,19,21 and 23). Note that all calculus for high entropy